\documentclass{article}

\usepackage[preprint]{neurips_2024}

\usepackage[utf8]{inputenc} % allow utf-8 input
\usepackage[T1]{fontenc}    % use 8-bit T1 fonts
\usepackage{hyperref}       % hyperlinks
\usepackage{url}            % simple URL typesetting
\usepackage{booktabs}       % professional-quality tables
\usepackage{amsfonts}       % blackboard math symbols
\usepackage{nicefrac}       % compact symbols for 1/2, etc.
\usepackage{microtype}      % microtypography
\usepackage{xcolor}         % colors
\usepackage{amsmath}
\usepackage{graphicx}       % graphicx for figures
\usepackage{enumitem}

\title{Continuous Speech Tokens Makes LLMs Robust Multi-Modality Learners}

\author{%
    \textbf{Ze Yuan}$^1$\thanks{Work done during an internship at Microsoft speech team.} \ \ 
    \textbf{Yanqing Liu}$^2$\thanks{Corresponding author: yanqliu@microsoft.com} \ \ 
    \textbf{Shujie Liu}$^2$ \ \
    \textbf{Sheng Zhao}$^2$ \\
    $^1$Tsinghua University\\
    $^2$Microsoft Corporation
}

\begin{document}

\maketitle

\begin{abstract}

Recent advances in GPT-4o like multi-modality models have demonstrated remarkable progress for direct speech-to-speech conversation, with real-time speech interaction experience and strong speech understanding ability. However, current research focuses on discrete speech tokens to align with discrete text tokens for language modelling, which depends on an audio codec with residual connections or independent group tokens, such a codec usually leverages large scale and diverse datasets training to ensure that the discrete speech codes have good representation for varied domain, noise, style data reconstruction as well as a well-designed codec quantizer and encoder-decoder architecture for discrete token language modelling. This paper introduces Flow-Omni, a continuous speech token based GPT-4o like model, capable of real-time speech interaction and low streaming latency. Specifically, (i) instead of cross-entropy loss only, we combine flow matching loss with a pretrained autoregressive LLM and a small MLP network to predict the probability distribution of the continuous-valued speech tokens from speech prompt. (ii) we incorporated the continuous speech tokens to Flow-Omni multi-modality training, thereby achieving robust speech-to-speech performance with discrete text tokens and continuous speech tokens together. Experiments demonstrate that, compared to discrete text and speech multi-modality training and its variants, the continuous speech tokens mitigate robustness issues by avoiding the inherent flaws of discrete speech code’s representation loss for LLM. See \url{https://cognitivespeech.github.io/flowomni} for demos of our work.

\end{abstract}

\begin{figure*}[h!]
    \centering
    \includegraphics[width=0.80\textwidth]{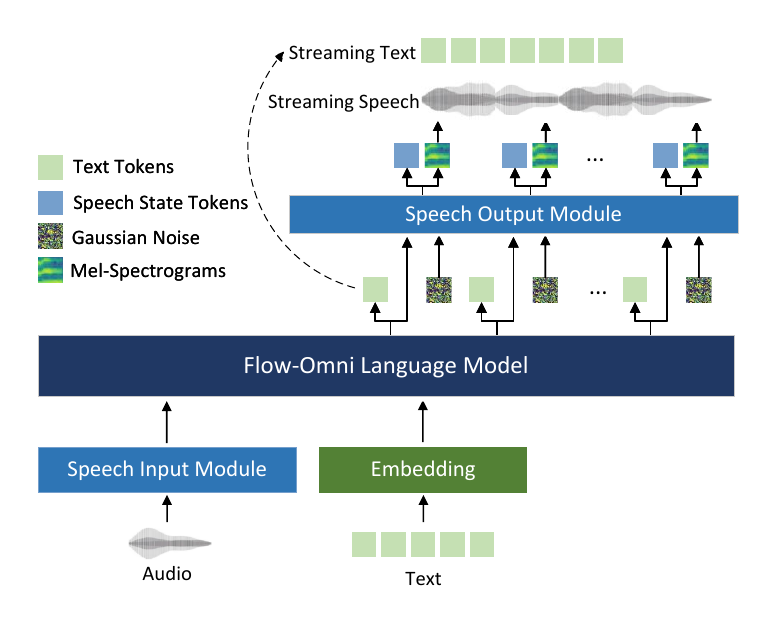}
    \vspace{-1.2mm}
    \caption{The overview of Flow-Omni. Unlike discrete tokens based language modeling approaches,  Flow-Omni generates the continuous mel-spectrogram conditioned on speech prompt autoregressively, using a single-stage decoder-only model its backbone, a whisper and adapter based model as speech understanding module, a light output adapter including flow matching MLP layers as speech generation module. 
    }
    \label{fig:overview}
\end{figure*}

\section{Introduction}

Cascaded speech interaction systems have powered the population from Microsoft Cortana to Apple Siri, Amazon Alexa in early age voice assistants, which first triggers an automatic speech recognition system that transcribes the user’s speech query, and then a natural language understanding model and a natural language generation model using template or model based solution compose the response text, finally, a text-to-speech system convert text to waveform for end users. The latency and the interaction experience of such application is not ideal for answer relevance, emotion conveying quality and yet latency. Recently, multi-modality large language models (LLM) like GPT-4o \citep{hurst2024gpt} have demonstrated better speech interaction experience and lower streaming latency by integrating speech recognition, language understanding, language generation and speech generation all in one to optimize.

Generally, there are two primary approaches for multi-modality language models. The first approach utilizes a pretrained decoder-only LLM and freezes the LLM parameters, optimizing speech or image adapters placed either before or after the decoder. This modular design simplifies integration with existing LLMs while preserving their pretrained knowledge. The second approach involves fine-tuning the LLM on multimodal datasets directly, without employing adapters. While this method offers tighter integration across modalities, it requires extensive computational resources and large-scale multimodal datasets to avoid overfitting. The second category finetunes the LLM with all multi-modality dataset without adapters. \citep{dao2024ichigo} is capable of generating and reasoning with mixed sequences of arbitrarily interleaved textual and speech content, expanding on tasks like understanding speech and text-only language models. \citep{ji2024wavchat,zhang2024omniflatten,defossez2024moshi,wang2024freeze, chen2024emova,cui2024recent,fang2024llama, ji2024wavchat} use multi-stage progressive post-training scheme to adapt a text-based large language model backbone into a robust speech-text dialogue model by first conducting modality alignment, and then performing dialogue learning. \citep{xie2024mini,xie2024mini2} propose a parallel text and audio generation method that leverages minimal additional data and modules to rapidly transfer a language model’s text capabilities to the audio and image modality. Additionally, their latest version employs a single model to simulate the visual, speech, and textual capabilities of GPT-4o.

Although multimodal language models built upon discrete text and discrete speech tokens have achieved remarkable progress in recent years, their end-to-end quality remains far from satisfactory due to several inherent limitations. First, discrete speech tokens are derived originally from audio codec that has representation loss especially for samples not seen in training, such as in high pitch, noise, emotion scenarios. Second, the discrete tokens highly depend on quantizer architecture to achieve high reconstruction quality, generally multi-layer residual vector quantization (VQ) or multi group parallel VQ is required, which introduces many VQ tokens at the same step or the total token number in one sequence is too long to model compared to text tokenizer algorithm like BPE. On the other hand, reducing the size of the codebook or adopting a single-layer codebook VQ approach to generate shorter speech token sequences poses significant challenges to maintaining high speech quality, resulting poor intelligibility and similarity. Moreover, these limitations in speech tokenization can have broader implications when the model is trained jointly with a text-based LLM. Inaccurate or low-fidelity speech tokens introduce noise into the training process, disrupting the alignment between speech and text modalities.  

In this work, to address the limitations associated with discrete speech tokens based multi-modality language models, we propose a continuous mel-spectrogram based multi-modality language model (Flow-Omni). Flow-Omni uses a pretrained LLM model which autoregressively predicts mel-spectrogram with flow matching module based on previous mel-spectrogram and speech prompt tokens, inherent the sampling ability like discrete tokens and avoid the flaws associated with inaccurate discrete codes. 

Specifically, given the speech prompt, Flow-Omni first encodes the waveform into latent embeddings with speech input module consisting of Whisper encoder and then projects it as autoregressive input to conditionally generate mel-spectrogram with speech output module consisting of flow matching predictor. The mel-spectrogram is finally converted into a waveform using a 24KHz vocoder. The training process combines the discrete cross-entropy loss and continuous vector regression loss of flow matching. We conducted evaluations of Flow-Omni on collected training datasets. Experimental results demonstrate that the proposed method is more robust than discrete speech token based system.

In summary, we make the following contributions. 
\begin{itemize}
\item We propose a multi-modality LLM with continuous speech tokens, which treats ASR (speech understanding) and TTS (speech generation) as different language model tasks with continuous mel-spectrogram as an intermediate representation to replace the popular discrete speech token from audio codec, the codec free pipeline simplifies LLM training and improves speech generation and robustness.
\item we verified flow matching based speech output module and discrete text token based LLM can be joint trained together and eventually worked as a more robust speech to speech model. 
 \end{itemize}

\section{Related work}

\subsection{Text-to-speech}

Text-to-speech (TTS) synthesis aims to generate natural speech from text with a high degree of similarity and intelligibility. Along with the progress of deep learning, significant improvements have been made in recent years from two stage model (text to acoustic model and acoustic to waveform vocoder) \citep{li2018close,li2019neural,li2020robutrans, li2020moboaligner,chen2021adaspeech,liu2021delightfultts,zhang2022mixed} to acoustic module and vocoder joint training models \citep{liu2022delightfultts,tan2024naturalspeech,kim2021conditional}, until recent large diffusion or language model based system \citep{wang2023neural,chen2024vall, xue2023foundationtts, zhang2023speak, shen2023naturalspeech, leng2023prompttts, kanda2024making, ju2024naturalspeech,zhang2024boosting, han2024vall, eskimez2024e2, meng2024autoregressive,li2024investigating} for zero shot TTS or single speaker finetuning TTS. Some systems, trained with clean single-speaker speech data recorded in professional studios, have even achieved human-level naturalness for limited domain generation \citep{liu2022delightfultts}. As a single task, both finetuning voices or zero shot TTS have been well explored for naturalness, similarity and expressiveness, however when combined with multi-modality large language model the speech representation and generation efficiency are yet to explore.

\subsection{Flow matching for generative modeling}

Flow matching \citep{lipman2022flow,tong2023improving,liu2022flow} learns a time dependent vector field that transports between two probability distributions for training continuous-time generative models, which is able to present an efficient approach to learn ordinary differential equations (ODEs)ssssssss, using a simple vector-field regression loss called conditional flow matching (CFM), as an alternative to learning score functions for DPMs or using numerical ODE solvers \citep{shaul2023bespoke} at training time like classic CNFs. Crucially, by leveraging ideas from optimal transport \citep{khrulkov2022understanding}, CFM can be set up to yield ODEs that have simple vector fields that change little during the process of mapping samples from the source distribution onto the data distribution. Diffusion or flow matching based model in text-to-image generation \citep{rombach2022high}, have recently shown remarkable results and established new benchmarks for generating high-quality, photorealistic images from text prompts. Additionally, recent studies have also demonstrated the effectiveness of diffusion models in various other generation tasks, such as text to speech or music and text to video \citep{khachatryan2023text2video,schneider2023mo}

\subsection{Multi Modality LLMs}

The recent release of GPT-4o with advanced voice capabilities has sparked significant global interest in the development of LLM-based technologies for low-latency, emotional speech powered interactions. Early examples of excellence in this field include Spectron \citep{nachmani2023spoken} and SpeechGPT \citep{zhang2023speechgpt}, which introduced the A-T-T-A framework to enable end-to-end speech input and output capabilities. Models such as VITA \citep{fu2024vita} and Qwen-audio2 \citep{chu2024qwen2} further support speech input but rely on external TTS systems for speech synthesis, producing text outputs directly. Advancing beyond these approaches, Mini-Omni \citep{xie2024mini} introduced a parallel generation mechanism for text and audio, enabling the model to perform inference directly in the audio input and output. Similarly, Llama-Omni \citep{fang2024llama} and Moshi \citep{defossez2024moshi} adopted parallel processing techniques to further enhance multi modal capabilities. Additionally, LSLM \citep{ma2024language} and Moshi \citep{defossez2024moshi} explored full-duplex interaction by incorporating simultaneous listening and speaking signals as inputs.

\section{Flow-Omni}
\label{gen_inst}

The model architecture of Flow-Omni is illustrated in figure \ref{fig:overview}. We add pre and post modules for speech understanding and generation upon a pretrained LLM, specifically the post generation module \citep{fan2024fluid} is designed to predict continuous mel frames, replacing the traditional discrete speech token from audio codec quantizers. This modification eliminates the need for separate pretrained and well-designed audio codec, thereby improving the robustness of the speech reconstruction and multi-modality capabilities. In this section, we provide a detailed discussion of the Flow-Omni architecture, the speech output module for continuous speech tokens, and the training strategy that leverages both flow matching loss and LLM cross entropy loss.

\subsection{Architecture}

\paragraph{Speech input module}

Flow-Omni employs the Whisper \citep{radford2023robust} encoder as its speech encoder \citep{xie2024mini}. The Whisper model features a highly robust architecture, comprising an audio encoder designed to transform raw audio waveforms into compact, high-dimensional continuous speech embeddings. This architecture is particularly adept at handling speech inputs across diverse dimensions, including various languages, accents, prosodies, and background noise levels. Additionally, Whisper’s pre training on an extensive and diverse dataset endows it with strong generalization capabilities, enabling it to effectively capture and encode speech features, even in low-resource or unseen conditions. Following the whisper encoder, several linear blocks are used to further project the speech features to LLM hidden dimensions acting as speech input tokens for LLM training.

\paragraph{Pretrained language model}

Flow-Omni integrates Qwen2 \citep{yang2024qwen2} as the LLM decoder for Chinese dataset experiments. Qwen2 is a state-of-the-art language model optimized for superior performance in both understanding and generating language, which consists of a transformer-based decoder with self-attention mechanisms, allowing it to efficiently capture complex linguistic structures. These capabilities make Qwen2 highly effective for downstream tasks requiring detailed comprehension and precise language generation. Furthermore, its inherent compatibility with multimodal frameworks facilitates seamless integration with the Whisper encoder. This integration enhances the overall system’s efficiency by bridging high-quality speech embeddings from Whisper with Qwen2’s advanced language processing capabilities. The pretrained model weight is first frozen for modality alignment then all the model parameters are finetuned to further improve generation and understanding quality.

\begin{figure}
  \centering
  \includegraphics[width=1.0\textwidth]{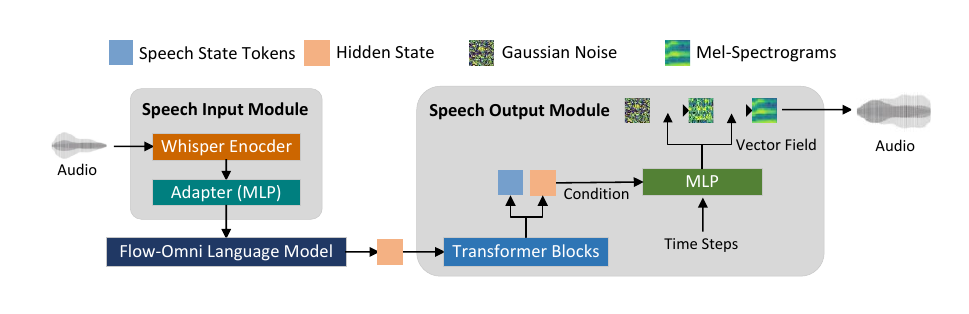}
  \caption{The architecture of speech input and output module. The feature of the input audio is first extracted by Whisper encoder and passed through the adapter to match the dimension of LLM. The LLM hidden is preprocessed by several transformer blocks and then a simple MLP based flow matching predictor is used to predict the mel vector with time step and LLM hidden as condition.}
   \label{fig:figure2}
\end{figure}

\paragraph{Speech output module}

The discrete token approach involves converting speech signals into a sequence of quantized ids, typically achieved via vector quantization or clustering methods. However, quantization introduces information loss, and the discrete nature of tokens can make it challenging to capture subtle variations in speech, like locale, prosody and style. Additionally, discrete representations may struggle with domain transfer, as the token vocabulary is tied to the training data distribution and may fail to generalize to unseen acoustic conditions or languages. Besides, the discrete token approach relies heavily on a high-quality pretrained codec as well as its deliberated design on codebook size and codebook number, etc. In contrast, the continuous token method bypasses the quantization step and directly predicts detailed acoustic features. This approach allows for the generation of highly realistic speech, as the model retains access to the full spectrum of acoustic information. Moreover, continuous token models are not limited by a fixed vocabulary, providing more flexibility, and can be seamlessly integrated with neural vocoders to further improve speech quality. Thus, inspired by masked autoregressive model (MAR) \citep{li2024autoregressive,wu2024janus,liu2024mardini}, we build a speech output module for predicting continuous-valued mel-spectrogram in Flow-Omni. The mel-spectrogram frame is obtained by modeling the vector field of conditional probability density paths. The detailed architecture of speech input and output modules are shown in figure \ref{fig:figure2}. It utilizes transformer blocks to obtain the hidden state from the output of language model as condition related to acoustic information, and predicts the state of the current speech stream, such as end, pad, and generating. The conditional vector field is modeled using a small MLP network conditioned on the output of transformer blocks, predicting the conditional probability density path to progressively denoise the random noise into the target mel-spectrogram. The current predicted mel-spectrogram token is then converted to waveforms by a GAN based vocoder \citep{kong2020hifi,kumar2019melgan}. The predicted mel-spectrogram is also fed input LLM decoder to autoregressively predict the next mel-spectrogram frame.

\subsection{Parallel sequence modeling}

\begin{figure}
  \centering
  \includegraphics[width=0.7\textwidth]{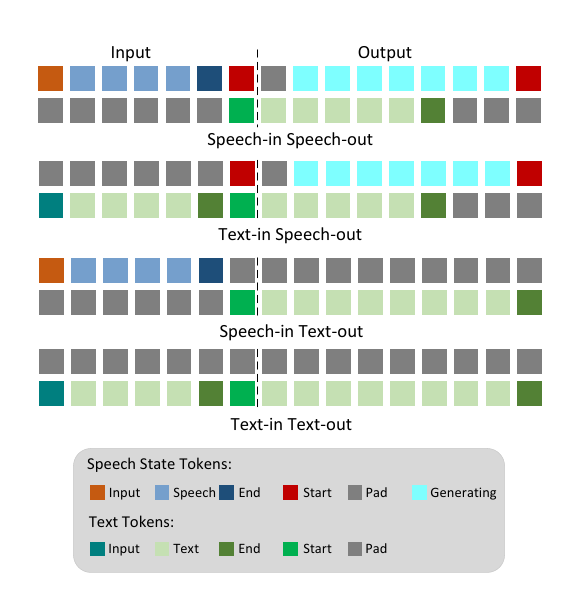}
  \caption{Parallel sequence of different tasks. When the input to the model is in either audio or text format, the other modality is filled with pad tokens. The text response follows the same format as in LLMs, and it is generated in parallel with the speech response, which is used to guide the synthesis of speech.}
     \label{fig:figure3}
\end{figure}

We extend the speech capability of language models by jointly modeling parallel speech sequence and text sequence \citep{xie2024mini,defossez2024moshi}. In Flow-Omni, we use only one speech sequence, which is used to represent the state of the speech, such as speech start, pad, stop, generating, while the mel-spectrogram is predicted by speech output module. The model can receive speech or text input and give speech or text response. There are four kinds of tasks in total, and their token sequence formulation are illustrated in figure \ref{fig:figure3}. Speech state sequence and text sequence are input and generated in parallel. The text input sequence is padded with pad tokens when the input is speech and vice versa. When the response is speech, the text tokens are also generated in parallel. To make full use of the text information, the speech delay generating is applied, and the first frame of the mel-spectrogram is generated one step later than the first text token. In speech inference, the mel-spectrogram is predicted using speech output module only when the speech state is “generating” and added to the autoregressive sequence.

\subsection{Training Objective}

Flow-Omni employs both language model cross entropy loss and conditional flow matching loss. The LM loss is used to train the ability of the LM to model the joint sequence. The conditional flow matching loss is used to train the speech output module's ability to model the conditional vector field of mel-spectrogram.

Given the ground-truth of the response sequence tokens of text \(y^t = (y_i^t)_{i \leq n}\) and speech state \(y^s = (y_i^s)_{i \leq n}\), and the estimated logits of text \(l^t = (l_i^t)_{i \leq n}\) and speech state \(l^s = (l_i^s)_{i \leq n}\) obtained from the model, the LM loss function is constructed with cross-entropy,

\begin{equation}
\mathcal{L}_{\text{LM}}(l^t,l^s,y^t,y^s) = \frac{1}{n}\sum_{i=1}^n\left(\text{CE}(l_i^t,y_i^t)+\text{CE}(l_i^s,y_i^s)\right)
\end{equation}

We use the approach of CFM to train the speech output module. Let \(x_1:\mathbb{R}^d\) denote the target mel-spectrogram, its distribution is \(q(x_1)\). The way to generate target mel-spectrogram from the data distribution \(q\) is to construct a probability density path \(p_t:[0, 1] \times \mathbb{R}^d \rightarrow \mathbb{R} > 0\), where \(t \in [0,1]\) and \(p_0(\mathbf{x}) = \mathcal{N}(x; \mathbf{0}, \mathbf{I})\), such that \(p_1(x)\) approximates the data distribution \(q(x)\). Define a learnable vector field \(v_t: [0,1] \times \mathbb{R}^d \rightarrow \mathbb{R}^d\), which generates the flow \(\boldsymbol{\phi}_t: [0,1] \times \mathbb{R}^d \rightarrow \mathbb{R}^d\) through the ODE

\begin{equation}
\frac{d}{dt} \boldsymbol{\phi}_t(x) = v_t(\boldsymbol{\phi}_t(x))
\end{equation}

Suppose the vector field \(u_t\) that can generate a probability path \(p_t\) from \(p_0\) to \(p_1 \approx q\). The flow matching loss is

\begin{equation}
\mathcal{L}_{\text{FM}}(\theta) = \mathbb{E}_{t, p_t(x)} \left\|u_t(x) - v_t(x; \theta) \right\|^2,
\end{equation}

where \(t \sim \mathcal{U}[0,1]\) and \(v_t(x; \theta)\) is modeled by a small MLP network with parameters \(\theta\) in Flow-Omni. Nevertheless, flow matching loss is intractable because it is non-trivial to get access to the vector field \(u_t\) and the target probability \(p_t\). In practice, conditional flow matching loss is used, and it has been proved that the optimization effects are consistent. conditional flow matching loss can be expressed with

\begin{equation}
\mathcal{L}_{\text{CFM}}(\theta) = \mathbb{E}_{t, q(x_1), p_t(x \mid x_1)} \left\| u_t(x \mid x_1) - v_t(x; \theta) \right\|^2.
\end{equation}

Finally, Flow-Omni is trained with loss function expressed

\begin{equation}
\mathcal{L} = \mathcal{L}_{\text{LM}}+\mathcal{L}_{\text{CFM}}
\end{equation}

\subsection{Inference sampling}

Discrete speech token based LLM can do sampling at each inference step with settings like top-P or temperature, in this way the output can be diverse in generated prosody. In Flow-Omni, sampling can be achieved by scaling flow matching temperature too. The conditional probability density path is generated by an optimal transport conditional vector field. Define the mean and standard deviation of the conditional probability path to change linearly in time. \(t:0\rightarrow 1 \), the probability density changes from \(p_0(x) = \mathcal{N}(x; \mathbf{0}, \mathbf{I})\) to  \(p_1(x\mid x_1)\), the mean and standard deviation can be expressed as

\begin{equation}
\mu_t(x)=tx_1;\quad \sigma_t(x)=1-(1-\sigma_{\min})t
\end{equation}

where \(\sigma_{\min}\) is a positive number close to zero. Thus, the conditional flow of this optimal transport is

\begin{equation}
\boldsymbol{\phi}_t(x) = (1 - (1 - \sigma_{\min})t)x + tx_1
\end{equation}

The vector field corresponding to the conditional flow is

\begin{equation}
u_t(x \mid x_1) = \frac{x_1-(1 - \sigma_{\min})x}{1 - (1 - \sigma_{\min})t}
\end{equation}

The MLP in the speech output module is used to predict \(u_t\). At inference time, assuming that the predicted value of \(u_t\) is \(v_t\), we can predict the mel-spectrogram \(x_1\) from the random noise \(x_0\), as illustrated by \citep{li2024autoregressive} sampling can be achieved by temperature control on random noise, and the calculation process can be expressed as

\begin{equation}
x_{t+\Delta t} = x_t + v_t \Delta t 
\end{equation}

\section{Experiment}

\label{headings}

\subsection{Datasets}

In our experiments, we used the Chinese speech-text dataset Aishell-1, WenetSpeech and the base version of dialogue corpus LCCC. 

\paragraph{Aishell-1 and WenetSpeech}
The total recording duration of the Aishell-1 dataset is 178 hours, recorded by 400 different speakers. This dataset contains a large number of recordings covering a variety of accents, speech speeds, and emotional variations. The WenetSpeech dataset contains 10,000+ hours of Mandarin speech data, all from YouTube and Podcast, covering a variety of fields and speaking styles, including audiobooks, live commentaries, documentaries, dramas, interviews, news, reading, discussion, variety shows and more. These data cover different Internet audio and video, noise background conditions. We trained the model with the transcripts and the mel-spectrograms extracted from the original audios. 
\paragraph{LCCC-base}
The LCCC-base dataset contains 6.8 million conversation rounds, derived from Weibo conversational data and open-source conversation datasets, and has undergone a rigorous cleaning process. We first removed repeated punctuations from the text and filter out Q\&A pairs less than five words in length and then use Azure TTS service \footnote{\url{https://learn.microsoft.com/en-us/azure/ai-services/speech-service/text-to-speech}} to synthesize the text in the data set into Q\&A speech. The question speech was synthesized into the voice of multiple speakers, and the answer speech was fixed into the voice of the same person, to obtain the Q\&A dataset with both text and speech. Then the mel-spectrograms were extracted from the audios for the training of the model.

\subsection{Training details}

We adopted a two-stage training strategy to make Flow-Omni extend the speech capability based on the pretrained language model.
\begin{itemize}
  \item In the first stage, we focused on modal alignment to ensure that the speech input and output modules can effectively extract audio features interpretable by the language model and synthesize speech outputs based on the language model. To achieve this, the parameters of the pretrained language model and the Whisper encoder were frozen, while the parameters of all other modules remained trainable. Training was conducted using the Aishell-1 and WenetSpeech datasets, with automatic speech recognition (ASR) and text-to-speech (TTS) serving as the training tasks. For the ASR task, the parallel sequence format aligned with the "speech-in text-output" depicted in figure \ref{fig:figure3}. Similarly, for the TTS task, the parallel sequence format followed the output section of "text-in speech-output" in figure \ref{fig:figure3}.
  \item In the second stage, we fine-tuned the entire model, allowing all parameters to be trainable. This stage leveraged the LCCC-base dataset supplemented with speech data and incorporates all four tasks outlined in figure \ref{fig:figure3}. The sequence formats were consistent with those specified in the figure. After this comprehensive training process, the resulting model can effectively understand and respond to both speech and text inputs, achieving seamless multimodal functionality.
\end{itemize}

During the training process, we utilized the adam optimizer in conjunction with a cosine annealing learning rate scheduler to ensure efficient convergence. The initial learning rate for both training stages was set to 1e-4, gradually reduced to a minimum of 5e-6 according to the annealing schedule. All experiments were conducted using NVIDIA V100 GPUs.

\subsection{Model configuration}

\paragraph{Speech input module}

For the speech input module, audio features are extracted using the encoder of the Whisper-small model. The Whisper encoder first converts the raw audio into a mel-spectrogram. The mel-spectrogram is then passed through a two-layer CNN to capture local features of the audio signal. Subsequently, the features are fed into a 12-block transformer encoder, which effectively extracts global dependencies and contextual information from the audio. The resulting hidden states are then processed through a two-layer MLP to align the output dimension with the language model. This architecture ensures robust feature extraction and seamless integration with the language model for subsequent processing tasks.

\paragraph{Pretrained language model}

Flow-Omni can use any language model as the base model, and other modules can be seamlessly integrated. In this study, we use Qwen2-0.5B as the language base model, which is a small language model with 24 transformer blocks and model dimension of 896, using the grouped query attention.

\paragraph{Speech output module}

The Speech output module contains 6 transformer blocks that continue to extract acoustical hidden from the hidden states of the language model output and predict the current speech state. A 3-layer MLP uses the output of the transformer blocks as the condition to predict the conditional vector field for generating the target mel-spectrogram, together with timestep embeddings.

\paragraph{Vocoder}

The vocoder adopts the same architecture as the generator in \citep{liu2022delightfultts}, conditioned by mel-spectrograms. Adversarial training combines both multi-scale and multi-period discriminators as \citep{liu2022delightfultts}, the training set is Libritts \citep{zen2019libritts} for speaker independent reconstruction.

\subsection{Results}

We evaluated the quality of speech responses generated by Flow-Omni model. We use Azure TTS service to synthesize the texts in the test set of LCCC-base into speech and input them into Flow-Omni model. The speech responses and corresponding texts generated by the model are used to compute WER. For comparison, we used the same Chinese dataset, trained a Mini-Omni model with the same base language model, and calculated WER in the same way. The results are presented in table \ref{tab:table1}, indicating that the application of continuous-value tokens is effective to improve the quality of speech.

\begin{table}[h]
\centering
\setlength{\tabcolsep}{20pt} 
\renewcommand{\arraystretch}{1.5} 
\begin{tabular}{ccc} 
\toprule
\textbf{Model} & \textbf{WER} \\ \midrule
Mini-Omni & 10.84  \\
Flow-Omni & 8.81 \\ \bottomrule
\end{tabular}
\vspace{10pt}
\caption{WER results of generated speech for Mini-Omni and Flow-Omni.}
\label{tab:table1}
\end{table}

\section{Conclusion}

In this study, we propose a continuous speech representation-based multi-modality language modeling approach for speech-to-speech QA tasks which eliminate the use of discrete vector quantization from well-designed quantizer in codec. By exploring the potential of mel-spectrograms within the paradigm of language modeling, the proposed Flow-Omni directly predicts mel-spectrograms conditioned on speech prompt with flow matching. With the aid of flow matching vector field regression loss, Flow-Omni is capable of producing more diverse and robust speech response.

%\section*{References}
%\newpage
%\bibliographystyle{rusnat}
%\bibliography{mybib}

\bibliography{mybib}
\bibliographystyle{plainnat}

%%%%%%%%%%%%%%%%%%%%%%%%%%%%%%%%%%%%%%%%%%%%%%%%%%%%%%%%%%%%

%%% END INSTRUCTIONS %%%

\end{document}